\documentclass[prb,twocolumn,showpacs,preprintnumbers,amsmath,amssymb]{revtex4}

\usepackage{amsfonts}
\usepackage{latexsym}
\usepackage{graphicx}

\newcommand{\Eqref}[1]{Eq.~(\ref{#1})}

\newcommand{\nn}{\nonumber}
\newcommand{\be}{\begin{equation}}
\newcommand{\ee}{\end{equation}}
\newcommand{\bear}{\begin{eqnarray}}
\newcommand{\ear}{\end{eqnarray}}

\begin{document}

\title{Burgulence and Alfv\'en waves heating
mechanism of solar corona}

\author{T.M.~Mishonov}
\email[E-mail: ]{mishonov@phys.uni-sofia.bg}
\author{Y.G.~Maneva}
\email[E-mail: ]{yanamaneva@gmail.com}
\affiliation{Department of Theoretical Physics, Faculty of Physics,\\
University of Sofia St.~Kliment Ohridski,\\
5 J. Bourchier Boulevard, BG-1164 Sofia, Bulgaria}

\date{\today}

\begin{abstract}
Heating of magnetized turbulent plasma is calculated in the
framework of Burgers turbulence [A.M. Polyakov, Phys. Rev. E.
\textbf{52}, 6183 (1995)]. Explicit formula for the energy flux of
Alfv\'en waves along the magnetic field is presented. The Alfv\'en
waves are considered as intermediary between the turbulent energy
and the heat. The derived results are related to a wave channel of
heating of the solar corona. If we incorporate amplification of
Alfv\'en waves by shear flow the suggested model of heating can be
applied to analysis of the missing viscosity of accretion discs and
to reveal why the quasars are the most powerful sources of light in
the universe. We suppose that the Langevin-Burgers approach to
turbulence we have applied in the current work can be also helpful
for other systems where we have intensive interaction between a
stochastic turbulent system and waves and can be used in many
multidisciplinary researches in hydrodynamics and MHD.
\end{abstract}

\maketitle

\section{Introduction}

For more than 60 years we are still facing the perplexing
astrophysical problem why the temperature of the solar corona is two
orders of magnitude higher than the temperature of the photosphere;
for review and references see for example
Ref.~[\onlinecite{Mandrini:00}, sec.~5.1]. The purpose of the
current work is to investigate the wave mechanism of heating of the
solar corona, according to which the energy is being transported by
magneto-hydrodynamical waves, generated in the stochastic convective
region. In the framework of this scenario the famous correlation
between the solar activity and the coronal emission in the X-ray
range can be easily explained. The novelty in the model is the
incorporation of the Burgers approach to the turbulence in the
problem for the calculation of the spectral density of the Alfv\'en
waves. In the framework of this model we derived an explicit
formulas for: 1)~the spectral density of the MHD waves, 2)~the total
energy density and 3)~the dissipated power per unit volume. In
perspective the results derived here can be generalized for a
realistic model of the turbulence and the distribution of density,
temperature and magnetic field. In the current work the model task
for calculation of the wave power, emitted by a turbulent half-space
and transmitted in a non-moving magnetized plasma, is considered as
an illustrative example. The homogeneous magnetic field is oriented
perpendicularly to the interface of both half-spaces.


\section{Burgulence}

The Burgers approach to the turbulence gives an opportunity for
approximate treatment of plenty of problems; as a tutorial lecture
on Burgulence and a source of main references we recommend
Ref.~[\onlinecite{LesHousches:00}]. This approach to the turbulence
has been introduced in the astrophysics by
Zeldovich\cite{Zeldovich:7082}, but didn't become very popular. Yet
in the last decade the incorporation of the quantum field theories
gave incitement to the theory of turbulence.
Kolmogorov\cite{Kolmogorov:41} power laws have been derived and
gradually the Burgers approach converted from a sophisticated high
energy physics problem\cite{Polyakov:95} to a standard tool for
investigation of the turbulence over different physical phenomena.
The concept of random driver in the special case of formation of
spicules and the problem of heating of space plasma in general was
considered in the recent works Ref. [\onlinecite{Erdelyi:04}]; here
we would like to emphasize on the similarity between propagation of
Alfv\'en waves in an infinite magnetized plasma and kink waves in
solar spicules\cite{Timuri:06}; for other papers devoted on the
subject see Ref. [\onlinecite{Goossens:00}]. There exist also
different ways of generation of waves by turbulence where the
influence of the random driver is negligible\cite{Hristov:9803}.

In the next section we will consider the application of the Burgers
approach to the theory of creation of Alfv\'en waves from the
turbulence and the stochastic granulation. In the $\beta \sim 1$
region of the solar atmosphere Alfv\'en waves can also be generated
by resonant conversion of acoustic oscillations\cite{Timuri:05}.

\section{Wave amplitudes and spectral densities}

According to the Burgers idea\cite{Burgers:48} the influence of the
turbulent vortices on the fluid can be modeled by introduction of a
random volume density of an external force $\mathbf F(t,\mathbf r)$
on the right hand side of the Navier-Stokes equation
\be
\label{Navier-Stokes}
\rho (\partial_t  + \mathbf{V} \cdot \nabla)\mathbf{V}
= - \nabla p + \eta \Delta \mathbf{V}
  + \mathbf{j} \times \mathbf{B}
  + \mathbf{F}. \ee
Analyzing low frequency and long wavelength phenomena one
can approximately suppose the random force to be a white noise
with a $\delta$-function correlator
\cite{Burgers:48,Polyakov:95,Zeldovich:7082,LesHousches:00}
\be
\label{WhiteNoiseCorrelator}
\langle
\mathbf{F}(t_1,\mathbf{r}_1)\mathbf{F}(t_2,\mathbf{r}_2)\rangle
=\tilde\Gamma{\rho}^2\delta(t_1-t_2)\delta(\mathbf{r}_1-\mathbf{r}_2)\openone,
\ee
where $\tilde\Gamma$ is the Burgers parameter. For the other
notions we use standard notations: $p$ is the pressure,
$\eta$ is the viscosity, $\mathbf B$ is the magnetic field.

The current $\mathbf{j}$ is given by Ohm's law
\be \label{Ohm2} \mathbf{j} =\sigma \mathbf E',\quad \mathbf
E'=\mathbf{E} + \mathbf{V}\! \times \mathbf{B} \, , \ee
where $\sigma$ is the electrical conductivity and $\mathbf E'$
expresses the effective electric field acting on the fluid.

In quasi-stationary approximation, when the $\partial_t \mathbf E$
term is negligible, the evolution of the magnetic field in a plasma
with constant resistivity is
\bear
\partial_t \mathbf{B} &=& \mathrm{rot}\,(\mathbf{V} \times \mathbf{B}) -
\nu_\mathrm m\mathrm{rot}\,\mathrm{rot}\,\mathbf{B}.
\ear
We consider an incompressible fluid
\be
\mathrm{div} \mathbf{V} = 0
\ee
with wave amplitudes for the velocity
\be
\mathbf V = (V_x, V_y, V_z),\quad V \ll V_{\mathrm A}
\ee
and the magnetic field
\be
\mathbf{B} = \mathbf{B}_0 + \mathbf{B'},\quad
\mathbf{B'}(t,\mathbf{r}) = (B'_x,B'_y,B'_z),\quad B' \ll B_0
\ee
being small compared to the Alfv\'en speed and the external magnetic
field. The $z$-axes is chosen in a vertical direction along the
constant magnetic field
\be
\mathbf{B}_0 = (0, 0, B_0)= B_0\mathbf e_z.
\ee
Analogously for the pressure we also suppose small deviations from a
constant value $p_0$
\be
p = p_0 + p'.
\ee
The linearized system of magnetohydrodynamic (MHD) equations then
reads
\begin{eqnarray}
\label{Euler}
&&\nn\partial_t\mathbf{V}
\!\!=\!-\frac{\nabla p'}{\rho}
\!+\frac{\mathbf{F}}{\rho}
\!+ \!\frac{B_0}{\mu_0 \rho}\! \left(
\begin{matrix}\partial_z B'_x -\partial_x B'_z\\
\partial_z B'_y  - \partial_y B'_z\\
0
\end{matrix}
\right)\!\!\!+ \nu_\mathrm k \Delta\mathbf{V}\!,\\
&&\nn\partial_t \mathbf{B'}\!
= \!B_0 \partial_z \mathbf{V} + \nu_\mathrm m \Delta \mathbf{B'},\\
&&\mathrm{div}\mathbf{B'}=0,\quad\mathrm{div}\mathbf{V}\!=0,
\end{eqnarray}
where
\be
\nu_m \equiv \frac{\varepsilon_0c^2}{\sigma},
\quad \nu_k \equiv \frac{\eta}{\rho}
\ee
are respectively the magnetic and kinematic viscosities. Here we
would like to emphasize that the incompressibility condition
$\mathrm{div}\mathbf V=0$ is applicable in the convective zone of
the sun, where the sound speed significantly exceeds the Alfv\'en
speed $c_\mathrm s \gg V_\mathrm A$. In this region the convective
circulation is treated as the influence of the turbulence in the
Burgers approach. The excited Alfv\'en waves then propagate along
the magnetic filed lines and reach the corona where the direction of
the inequality could be opposite.

Let us consider the evolution of the amplitudes of standing plane
waves
\begin{eqnarray}
\label{V_plane} \mathbf{V}(t,\mathbf{r})=V_A\vec{\upsilon_k}(t)
\sin(\mathbf{k}\cdot\mathbf{r}),\\
\label{B_plane} \mathbf{B'}(t,\mathbf{r})=B_0\mathbf{b_k}(t)
\cos(\mathbf{k}\cdot\mathbf{r}),\\
\nn \mathbf F(t,\mathbf r) =\rho V_\mathrm A \mathbf f_k(t)
\sin(\mathbf{k}\cdot\mathbf{r}),\\
 p' = \rho V_\mathrm A p_{_k}
\cos(\mathbf{k}\cdot\mathbf{r}),\\
\label{obratencorelator}
\langle
\mathbf{f^*_\mathbf{p}}(t_1)\mathbf{f_\mathbf{k}}(t_2)\rangle
=\Gamma\delta(t_1-t_2)
\delta_{\mathbf{p},\mathbf k}\openone.
\end{eqnarray}
This is a technical tool for evaluation of the average energy
density used in many textbooks on statistical physics. Due to the
negligible dependance on the boundary conditions usage of running
waves would lead to the same result.

Substitution of this ansatz in \Eqref{Euler} gives a separation of
the variables and for the wave amplitudes we have a system of
ordinary differential equations
\bear
\label{EulerFourier}
\mathrm d_t \vec\upsilon = p \,\mathbf k + f -
V_\mathrm A \left(
\begin{matrix}k_z b_x - k_x b_z\\
k_z b_y  - k_y b_z\\
0
\end{matrix}
\right) - \nu_\mathrm k k^2 \vec\upsilon,\\
\mathrm d_t\mathbf b = V_\mathrm A k_z\vec\upsilon - \nu_\mathrm m k^2 \mathbf b,\\
\mathbf k\cdot \mathbf b =0,\\
\label{incompressF}\mathbf k\cdot \vec\upsilon=0,
\ear
where for the sake of brevity the wave-vector indices are omitted.
Time differentiation of the incompressibility equation
\Eqref{incompressF} and substitution in \Eqref{EulerFourier} gives
the explicit form of the pressure
\be
\label{roga4}
 p = - \frac{\mathbf k \cdot \mathbf f}{k^2} -
V_\mathrm A b_z
\ee
Further the substitution of the pressure from Eq.~(\ref{roga4}) in
\Eqref{EulerFourier} gives
\bear
\label{startsyst}
\nn\dot {\vec \upsilon} &=& - V_\mathrm A k_z \mathbf b
- \nu_\mathrm kk^2 \vec \upsilon
+ \mathbf f_\perp,\\
\dot{\mathbf{b}}&=& V_\mathrm A k_z \vec \upsilon - \nu_\mathrm m k^2\mathbf b,
\ear
where
\be
\mathbf f_\perp = \hat\Pi_\perp\cdot\mathbf f
\ee
is the transverse projection of the external force with respect to
the wave-vector $\mathbf k$ by the polarization operator $\Pi_\perp$
\be
\label{polaroper}
\hat\Pi_\perp = \openone - \frac{\mathbf k\otimes\mathbf k}{k^2}= \left(
\begin{matrix}\frac{k_y^2+k_z^2}{k^2} \frac{-k_xk_y}{k^2} \frac{-k_yk_z}{k^2}\\
\frac{k_x^2+k_z^2}{k^2} \frac{-k_xk_y}{k^2} \frac{-k_yk_z}{k^2}\\
\frac{k_x^2+k_y^2}{k^2} \frac{-k_xk_z}{k^2} \frac{-k_yk_z}{k^2}
\end{matrix} \right).
\ee
This method for elimination of the pressure is used in Ref.
[\onlinecite{Chagelishvili:93}] dedicated on amplification of
Alfv\'en waves in the presence of shear flow. A continuation of this
research in the axial symmetric case is presented in Ref.
[\onlinecite{Rogava:03}].

 Let us investigate the eigen-modes $\propto \exp (-i\omega t)$
discarding for a moment the influence of the external force $\mathbf
f$. For the case of small dissipation the secular equations for all
the components are the same
\be
\left|\begin{matrix} -i\omega + \nu_\mathrm k k^2&& -V_\mathrm A k_z\\
V_\mathrm A k_z &&-i\omega + \nu_\mathrm m k^2\end{matrix}\right|
=0, \ee
with eigen-values
\be
\omega \approx \omega_\mathrm A - i\frac{\gamma}{2},
\ee
where $\omega_\mathrm A = V_\mathrm A k_z$ is the frequency of the
Alfv\'en waves and
\be
\label{attencoeff}
\gamma = \nu k^2,\quad \nu \equiv \nu_m + \nu_k
\ee
is the attenuation coefficient.

In regime of negligible magnetic viscosity the MHD system \Eqref{startsyst}
turns into the second order ordinary differential equation
\be \ddot{\vec\upsilon} = \nu_\mathrm k k^2 \dot{\vec\upsilon} -
\omega^2_\mathrm A\vec\upsilon + \dot{\mathbf f_\perp}. \ee
If we introduce the $x$-component of the displacement
\be
x(t) = \int_{0}^t \upsilon_x(t')\mathrm dt'
\ee
we obtain an effective oscillator equation under an external force
\be \ddot{x} = \nu_\mathrm k k^2 \dot{x} - \omega^2_\mathrm A x +
\mathbf f_\perp, \ee
which in case of negligible kinematic viscosity has an effective energy
\be
\mathcal{E}=\frac{1}{2}\left(\dot{x}^2+
\omega^2_\mathrm Ax^2\right).
\ee
This is the non-perturbed by the friction and the weak random noise
energy density. A harmonic oscillator under an external force is a
well-known mechanical problem, see Ref.~[\onlinecite{LandauI}],
Chap.~V, Eq.~(22,11). When the external force is a white noise a
simple averaging gives that the energy is a linear function of time;
in other words the random noise gives a constant power to the
oscillator
\be
\label{oscpower}
\mathrm d_t \langle \mathcal{E}\rangle =
\mathrm d_t \left\langle\frac{1}{2}\dot{x}^2+ \frac{1}{2}
\omega^2_\mathrm A x^2\right\rangle =
\mathrm d_t\left\langle\upsilon_x^{\,2}\right\rangle
= \frac{\Gamma}{2},
\ee
where
\be
\label{gama-gamatilde}
\Gamma \equiv \frac{\tilde\Gamma}{\mathcal V V^2_\mathrm A}
\ee
is the Burgers parameter in the correlator for the plane-wave
amplitudes of the transverse projection of the random force in
\Eqref{obratencorelator}
\be \langle
\mathbf{f^*_{\mathbf{p},\perp}}(t_1)\mathbf{f_{\mathbf{k},\perp}}(t_2)
\rangle =\Gamma\delta(t_1-t_2)\delta_{\mathbf{p},\mathbf k}\openone;
\ee
the technical details of this simple derivation will be omitted. Here
we suppose that the wave energy does not increase significantly
over one period
\be
\Gamma \ll \omega_\mathrm A
\ee
and the brackets stand for a period and noise averaging. In static
regime the condition for dynamic equilibrium requires that the power
received from the noise \Eqref{oscpower} has to be equal to the
dissipated power
\be
\mathrm d_t \langle \mathcal{E}\rangle
= - \gamma \langle \mathcal{E}\rangle
\ee
and this gives the dimensionless static averaged energy of the oscillator
\be
\label{static}
\mathcal{E}_\mathrm{st} = \frac{\Gamma}{2\gamma}
= \langle \upsilon_x^{\,2}\rangle_{\mathrm{st}}.
\ee

Now we are going to use this model example for derivation of the
spectral density of Alfv\'en waves. For the volume density of the
average wave energy we have
\be
\label{waveenergy}
 \langle E \rangle= \frac{\left\langle B'^2\right\rangle}{2\mu_0}
+ \frac{\left\langle V^2\right\rangle}{2\rho}.
\ee
This result may be rewritten as
\be
\label{preliminary}
\langle E \rangle= \frac{B^2_0}{4\mu_0}\langle b^2 \rangle + \frac{V^2_\mathrm A}{4\rho}
\langle\vec\upsilon^{\,2}\rangle = \frac{1}{2}\langle\vec\upsilon^{\,2}\rangle p_{_B},
\ee
where we have preformed a volume averaging
\be
\left\langle \sin^2(\mathbf k\cdot\mathbf r)\right\rangle = \frac{1}{2},
\ee
and used the definitions for the magnetic pressure
\be
p_{_B} \equiv \frac{B_0^2}{2\mu_0}
\ee
and Alfv\'en speed
\be
\frac{V^2_\mathrm A}{2\rho}=\frac{B^2_0}{2\mu_0}.
\ee
In \Eqref{preliminary} we can substitute the static squared
amplitude $ \langle \upsilon_x^{\,2}\rangle_{\mathrm{st}}$ from
\Eqref{static} and the attenuation coefficient from
\Eqref{attencoeff}. Taking into account the $y$ and $z$-components
of the Alfv\'en wave velocities results in an additional
triplication of the energy. Then for the static value of the density of the
wave energy we finally derive
\be
\label{SpectralDensity}
E_k\equiv  \langle E \rangle_{\mathrm{st}}
 = \frac{3}{4}\frac{\Gamma p_{_B}}{\nu k^2}.
\ee
This spectral density is the main detail for the statistical
analysis made in the next section.

\section{Energy density, heating rate and energy flux}

In order to calculate the total static energy density we have to
perform a summation of the spectral density \Eqref{SpectralDensity}
over all wave-vectors
\be
\label{eqfortotenergy}
E_\mathrm{tot} = \sum_k E_k =\mathcal V\int\frac{\mathrm
d^3\mathbf k}{{(2\pi)}^3}E_k.
\ee
The wave-vectors' cut-off is determined by the condition for the
existence of Alfv\'en waves, i.e. by the region where the Alfv\'en
frequency $\omega_\mathrm A$ equals the attenuation coefficient
$\gamma$
\be
V_\mathrm A k_z=\nu k^2.
\ee
This equation sets the natural cut-off $k_c$ for the vertical component
of the wave-vector $k_z$, parallel to the constant magnetic field
$\mathbf B_0$
\be
\label{cut-off}
k_c =\frac{V_\mathrm A}{\nu}.
\ee
Taking into consideration that the Alfv\'en waves are spreading
axially symmetric as they follow the magnetic field lines
we may rewrite the wave-vector as
\be
k^2 = k_z^2 + k_\rho^2,
\ee
which fixes the maximal value of the plane components of the
wave-vector $k_\rho$ as a function of the vertical component
cut-off $k_c$
\be
\label{maxwave-vec}
k_\rho^\mathrm{(max)}= \sqrt{k_z(k_c - k_z)}.
\ee
These cut-offs will be used as boundary conditions
in the calculations of the total wave energy, integrated
over all possible wave-vectors in \Eqref{eqfortotenergy},
and the corresponding wave power. In this way
considering the axial symmetry we obtain
\be
E_\mathrm{tot} = \frac{3\mathcal V\Gamma p_{_B}}{{(2\pi)}^3\nu}
\int_{k_z=\frac{1}{R}}^{k_c} \int_{k_\rho=0}^{k_\rho^\mathrm{(max)}}
\frac{1}{k_z^2+k_\rho^2}\mathrm d(\pi k^2_\rho)\mathrm dk_z,
\ee
where we have taken into account that due to infrared divergencies
the vertical component of the wave-vector should also have a minimal value,
determined by the typical size $R$ of the given magnetic field loop.
Then, having in mind \Eqref{gama-gamatilde}, the final expression
for the total energy becomes
\be
\label{totenergy}
E_\mathrm{tot} = \frac{3\tilde\Gamma
Rp_{_B}}{8\pi^2\nu^3}
 \left[
   \ln\left(\frac{V_\mathrm AR}{\nu}\right)-1
 \right].
\ee
Hereby, for the Alfv\'en waves energy flux along the magnetic
field we find
\be
\label{wavecurrent}
S =\frac{1}{2}E_{\mathrm{tot}}V_{\mathrm A}=
\frac{3\tilde\Gamma V_\mathrm A
Rp_{_B}}{16\pi^2\nu^3}
 \left[
    \ln\left(\frac{V_\mathrm AR}{\nu}\right)-1
 \right].
\ee
Multiplication of the spectral density with decay $\gamma$ gives the
heating rate and summation in the wave-vectors' space gives the
dissipated by the Alfv\'en waves per unit volume power
\be
\label{totalpower}
Q =\mathcal V\int E_k\gamma_k \frac{\mathrm d^3\mathbf k}{{(2\pi)}^3}
=\frac{\tilde\Gamma V_{\mathrm A}}{8\pi^2\nu^3}p_{_B}.
\ee
Perhaps magnetic field dependance of the heating rate can explain
the correlation between the magnetic flux and X-ray
luminosity\cite{Fisher:98}. In order to apply this result we need to
know the correlation between the magnetic field and the radius of
the active region.

The parameter with dimension length $R$ which we introduced due to
the infrared divergencies participate in the final formulas for the
total energy density \Eqref{totenergy} and energy flux
\Eqref{wavecurrent}. It seems very plausible if this detail could be
helpful to reveal the scaling correlations of the length of coronal
loops and other observable quantities. It is premature, however, to
start a detailed discussion based on a model example. The purpose of
the present work is to demonstrate that the Burgers approach should
be used for realistic computer simulations in the future.

\section{Discussion and conclusions}

Up to now, the coronal heating problem has not a satisfactory
explanation not because of a lack of interest, but rather due to
difficulties concerning the simultaneous obtaining of reliable
observational data for all processes running on the solar surface.
For determination of the parameters of each model as a rule we are
forced to use indirect means. The model, which we now present on the
arena suffers the same disadvantage. For its confirmation it is
necessary to know: 1)~the distribution of magnetic field near the
sunspots, 2)~velocity-velocity correlator in horizontal direction
or, which is equivalent, the correlator of the small stochastic
perpendicular components of the magnetic field, etc. In short, apart
from a realistic model for the turbulence we also need a realistic
model for all other variables. This naturally supposes a broad
interdisciplinary collaboration, which involves the whole spectrum
of the solar physics, from the statistical MHD to the X-ray
luminosity of the solar corona. The confirmation of the suggested
model requires realistic Monte Carlo simulations, for which the
analytical calculations done in the current work
\Eqref{wavecurrent}, \Eqref{totenergy}, \Eqref{totalpower} are only
test examples indispensable for a more realistic treatment of the
problem. We propose an oversimplified, and therefore solvable,
model, which realistic generalization will hopefully lead to an
adequate theory for the solar corona heating.

A natural continuation of the current investigation is to take into
account the influence of the shear flow over the magnetized
turbulent plasma. Preliminary numerical calculations on the behavior
of MHD waves in a shear flow show a significant amplification of the
slow magneto-sonic mode. The waves energy amplification is actually
a transformation of the shear flow energy into wave energy which at
the end dissipates into heat. This dissipation originates an
effective viscosity in the shear flow and the current scenario for
heating of the solar corona may also happen to be the theory for the
missing viscosity in accretion disks. Understanding the origin of an
additional viscosity in a magnetized turbulent plasma is rather
important for the proper explanation of the shining mechanism of
quasars, as well as for the angular momentum transport during
formation of compact astrophysical objects.

The theory for the coronal heating takes into consideration the
influence of the stochastic random noise of the turbulence and the
granulation over the evolution of Alfv\'en waves. A similar
mathematical scheme is used in the condensed matter physics for
calculation of the fluctuational
superconductivity\cite{Mishonov-Maneva-Penev:06}. In the physics of
superconductors the wave field is the effective Ginzburg-Landau wave
function and the noise is produced by Langevin thermal fluctuations,
which come from the inhomogeneous term in the TDGL equation.
Naturally, the solid state scheme is meticulously developed and the
final analytical results can be used for fitting of repeatable
experimental data from current-voltage characteristics. The present
work represents a realization of a similar idea in the new
astrophysical plasma areal.

\acknowledgments
Discussions and support by D.~Damianov, R.~Erdelyi,
I.~Rousev and I. Zhelyazkov are highly appreciated.

\bibliographystyle{apsrev}

\bibliography{AccretingDisks}

\end{document}